\documentstyle[12pt]{article}

\author{Gashimzade F.M. and Babaev A.M.\\
Institute of Physics, Azerbaijan Academy of Sciences, \\
370143 Baku, Azerbaijan, e-mail: semic@lan.ab.az}
\title{Kane Oscillator
}
\input tcilatex

\QQQ{Language}{
American English
}

\begin{document}

\maketitle
\begin{abstract}
The energy spectrum and wave functions for Kane oscillator describing the
spectra of electrons, light hole and spin-orbit-splitting bands in a quantum
dots with harmonic lateral confinement is found.
\end{abstract}

\section{Introduction}

As known, for description of an energy spectrum of quantum dots is used or
model ''infinite potential barries'', or model of harmonic confinement
[1].It was established, that the model of parabolic potential is realistic
for describing enough not too large quantum dots. Therefore, within the
framework of this model for the standard law dispersion electrons were
considered a number of problem of physics of quantum dots, including quantum
crystallizations electrons in an external magnetic field [2].

The semiconductor compounds (InAs, GaAs, InSb etc.) on which are created
quantum dots now, have a complex energy spectrum described multiband
Hamiltonian. In particular, nonparabolicity of a spectrum is possible to
take into account within the framework of eight-bands model Kane's[3]. Such
approach was applied in work [3], however because of complexity of the
obtained equation, the analysis of its solution was carried out in
frameworks rather special approximation.

The mentioned above complexity of the equation arises at a standard way of
introduction of parabolic confinement potential through scalar potential. If
introduce the external potential in a minimal way with the substituting[4]: 
\begin{equation}
\label{1}\stackrel{\rightarrow }{p}\rightarrow \stackrel{\rightarrow }{p}%
-i\beta \lambda \stackrel{\rightarrow }{r} 
\end{equation}

then as shown in [4,5,6,7] for Dirac Hamiltonian the oscillator equation
with additional constant term is obtained , which originated from
spin-orbital coupling.

Here we have applied the above-stated approach to obtain oscillators
equation from the equations invariance under the rotational group [8,9],
which directly gives a system of equations for radial functions for any
number of considered bands.

The obtained equation we have call Kane oscillator, by as an analogy to
Dirac oscillator . For to obtain the Kane's spectrum from the system of
equations (18) [8] let us consider the values $($ $j=1/2,\tau $\ $=0),$for
conduction band,for the heavy-and light hole $(j=3/2,\tau =1)$,and for the
spin-orbital-splitting band $(j=1/2,\tau =1)$.The first index characterizes
the weight of a irreducible representation and a second one indicates the
subspace with the same weight. We have chosen indexes for the states which
clearly show that they are created from corresponding $s$, $p$ states. In
order to give a physical meaning to equations we consider the coefficients
coupling $s$, and $p$ states, correspondingly, $\tau =0$, and $\tau =1$ to
be nonzero, where $\tau $ is number of the subspace.

After substituting 
$$
\frac d{dr}\rightarrow \frac d{dr}+\lambda \beta r 
$$

\begin{equation}
\label{2} \beta =\left( -1\right) ^{\left( \tau +\frac 12-\left| j_z\right|
\right) }\delta _{ij} 
\end{equation}

(where $j_z-$ is magnetic quantum number, $\lambda -$ is parameter
characterizes a steepness of a well) system of the equations , including
also dispersionless band of heavy hole has the form[8,9]:

$$
\left\{ \frac{-ia}{2\left( E-E_g\right) }\left[ \frac d{dr}+\lambda r+\frac{%
1\mp \left( l_0+\frac 12\right) }r\right] f_3^{\mp }\right. - 
$$
\begin{equation}
\label{3}-\left. \frac{i\sqrt{2}b}{E-E_g}\left\{ \left[ \frac d{dr}+\lambda
r+\frac{5\pm \left( l_0+\frac 12\right) }{2r}\right] f_2^{\pm }+\frac \alpha
rf_1^{\mp }\right\} +f_0^{\mp }\right\} =0 
\end{equation}

\begin{equation}
\label{4} \frac{i\sqrt{2}b}{E}\frac{\alpha }{r}f_{0}^{\pm }-f_{1}^{\pm }=0 
\end{equation}

\begin{equation}
\label{5} -\frac{i\sqrt{2}b}{E}\left[ \frac{d}{dr}-\lambda r-\frac{1\pm
\left( l_{0}+\frac{1}{2}\right) }{2r}\right] f_{0}^{\pm }+f_{2}^{\pm }=0 
\end{equation}

\begin{equation}
\label{6} \frac{-ia}{2\left( E+\Delta \right) }\left[ \frac{d}{dr}-\lambda
r+ \frac{1\mp \left( l_{0}+\frac{1}{2}\right) }{r}\right] f_{0}^{\mp
}+f_{3}^{\pm }=0 
\end{equation}

$$
\alpha =\frac{\sqrt{3}}2\sqrt{\left( l_0-\frac 12\right) \left( l_0+\frac
32\right) } 
$$
Here the following designations are used:

\begin{equation}
\label{7} 
\begin{array}{c}
f_{0}^{\pm }=f_{
\frac{1}{2},\frac{1}{2},0}^{l_{0}}\pm f_{\frac{1}{2},-\frac{1}{2},0}^{l_{0}}
\\ f_{1}^{\pm }=f_{
\frac{3}{2},\frac{3}{2},1}^{l_{0}}\pm f_{\frac{3}{2},-\frac{3}{2},1}^{l_{0}}
\\ f_{2}^{\pm }=f_{
\frac{3}{2},\frac{1}{2},1}^{l_{0}}\pm f_{\frac{3}{2},-\frac{1}{2},1}^{l_{0}}
\\ f_{3}^{\pm }=f_{\frac{1}{2},\frac{1}{2},1}^{l_{0}}\pm f_{\frac{1}{2},- 
\frac{1}{2},1}^{l_{0}} 
\end{array}
\end{equation}

As well as:

\begin{equation}
\label{8} 
\begin{array}{ccccccc}
\frac{C_{1/2,1/2}^{0,1}}{i\chi }=\frac{ia}{E-E_g} &  & \frac{%
C_{1/2,1/2}^{1,0}}{i\chi }=\frac{ia}{E+\Delta } &  & \frac{C_{1/2,3/2}^{0,1} 
}{i\chi }=\frac{ib}{E-E_g} &  & \frac{C_{1/2,1/2}^{1,0}}{i\chi }=\frac{ib}E 
\end{array}
, 
\end{equation}
where $E_g$ is the energy of the bottom of conduction band, $\Delta $ is the
spin-orbit splitting energy, The parameters $a,b,$ are matrix elements of
coupling between the conduction and valence bands. The quantities like $%
C_{1/2,1/2}^{0,1}$, $f_{1/2,1/2,0}^{l_0}$ etc. and $\chi $ are determined in
Gelfand et al $[8]$. The system of equations (1)-(6) are rewritten so that
to separate the independent solutions (''even'' and ''odd'').

\section{The Energy Spectrum}

Substituting (4)-(6) in (3) we shall obtain:

$$
\left\{ \left[ \frac{\partial ^2}{\partial r^2}+\frac 2r\frac \partial
{\partial r}-\frac{\left( l_0+\frac 12\right) \left( l_0+\frac 12\pm
1\right) }{r^2}-\lambda ^2r^2-3\lambda \right] \left[ \frac{a^2}{%
4(E-E_g)(E+\Delta )}+\frac{2b^2}{E(E-E_{g)}}\right] +\right. 
$$
\begin{equation}
\label{9}\left. +1\pm \left[ \frac{a^2}{2(E-E_g)(E+\Delta )}-\frac{2b^2}{%
E(E-E_{g)}}\right] \lambda (l_0+\frac 12\pm 1)\right\} f_0^{\pm }=0 
\end{equation}

Energy spectrum and the corresponding eigenfunctions are given by:

\begin{equation}
\label{10} \varphi (E)=2\lambda (N+\frac 32) 
\end{equation}

\begin{equation}
\label{11}\varphi (E)=\frac{4E(E-E_g)(E+\Delta )}{a^2E+8b^2(E+\Delta )}\pm
\lambda \left( l_0+\frac 12\pm 1\right) \frac{2a^2E-8b^2(E+\Delta )}{%
a^2E+8b^2(E+\Delta )}-3\lambda 
\end{equation}

\begin{equation}
\label{12}f_{0,n}^{\pm }=A_{nl_0\pm \frac 12}r^{l_0\pm \frac 12}\exp \left(
- \frac{\lambda r^2}2\right) L_n^{l_0\pm \frac 12+\frac 12}\left( \lambda
r^2\right) 
\end{equation}

where $L_n^{l_0\pm \frac 12+\frac 12}\left( \lambda r^2\right) -$is an
associated Laguerra polynomial, $N=2n+l_0\pm \frac 12$,$n=0,1,2,...$is
principal quantum number. The normalization constants are:

\begin{equation}
\label{13}A_{nl_{0\pm \frac 12}}=\left[ \frac{2\lambda ^{l_0\pm \frac
12+\frac 32}n!}{\Gamma (n+l_0\pm \frac 12+\frac 32)}\right] ^{\frac 12}
\end{equation}

$$
\lambda =\frac{m_n\omega }\hbar 
$$

The parameters $a$ and $b$ are related to the effective mass as follows[11]:

$$
\frac{\hbar ^2}{2m_n}=\frac{2b^2}{E_g}+\frac 14\frac{a^2}{E_g+\Delta }; 
$$

\begin{equation}
\label{14}\frac{\hbar ^2}{2m_{lh}}=\frac{2b^2}{E_g}; 
\end{equation}

$$
\frac{\hbar ^2}{2m_{sh}}=\frac 14\frac{a^2}{E_g+\Delta }, 
$$

where $m_n,$ $m_{lh}$ and $m_{sh}$ are the effective masses of electron,
light hole and spin-orbit splitting hole, correspondingly.

Case $a=\frac 2{\sqrt{3}}P,b=\frac 1{\sqrt{3}}P$ correspond to one
parameters Kane's model and using (14) we find for

\begin{equation}
\label{15}\varphi (E)=\frac{2m_n}{\hbar ^2}\left( \frac{E(E-E_g)(E+\Delta )}{%
E_g(E_g+\Delta )}\frac{E_g+\frac 23\Delta }{E+\frac 23\Delta }\mp \frac{%
\frac 23\Delta }{E+\frac 23\Delta }\frac{\hbar \omega }2\left( l_0+\frac
12\pm 1\right) -3\frac{\hbar \omega }2\right) 
\end{equation}

Using (12) to (4),(5) and (6), we find $f_1^{\pm },f_2^{\pm }$ and $f_3^{\pm
}$ :

\begin{equation}
\label{16}f_1^{\pm }=\frac{\imath \sqrt{2}b}E\frac \alpha rf_{0,n}^{\pm } 
\end{equation}

$$
f_2^{\pm }=\frac{\imath \sqrt{2}b}E\left[ \left( \frac{4n-1+2l_0\pm 1\mp
\left( l_0+\frac 12\right) }{4r}-\lambda r\right) f_{0,n}^{\pm }-\right. 
$$

\begin{equation}
\label{17}\left. -\frac{\sqrt{n\left( n+\frac 12+l_0\pm \frac 12\right) }}%
rf_{0,n-1}^{\pm }\right] 
\end{equation}

\begin{equation}
\label{18}f_3^{\pm }=-\frac{\imath a}{E+\Delta }\left\{ \left( \frac{2n\mp
1+l_0\left( 1\mp 1\right) }{2r}-\lambda r\right) f_{0,n}^{\mp }-\right. 
\end{equation}
\begin{equation}
\label{19}\left. -\frac{\sqrt{n\left( n+\frac 12+l_0\mp \frac 12\right) }}%
rf_{0,n-1}^{\mp }\right\} 
\end{equation}

The equations (15) describe the spectrum of electrons, light and
spin-orbital splitting holes bands .

As well as in a case Dirac oscillators in ground state energy appears twice
more, than for isotropic oscillator of the standard Schrodinger equation .

The equation (15) can appear useful to the analysis of influence
nonparabolicity on a energy spectrum electrons in a quantum dots. Recently
this problem is considered in works [10,11] within the framework of model of
a infinite potential barrier. The advantage of the given approach consists
in simplicity of the analysis of analytical expressions, in comparison with
numerical accounts [10,11].

The authors are grateful to E. Jafarov who addressed their attention to
works on Dirac oscillators.

Reference:



1.N.E.Kaputkina, Y.E.Lozovik Fiz. Tver.tela v.40,11,1753-1759 (1998)

2.N.E.Kaputkina, Y.E.Lozovik Fiz. Tver.tela v.40,9,2134-2135 (1998)

3. \`{O}. Darnhofer, U.Rossller, Rhys. Rev. B47, 23, 16 020 (1993).

4. Janes P Crawford J. Math. Phys. v. 34, 10, p. 4428-4435 (1993)

5. J.Benitez, R. P. Martinez y Romero Phys. Rev Lett. V.64, 14 (1990)

6. M. Moshinsky and A .Szezepanik, J. Phys. A 22, L817 (1989)

7. P. A. Cook Lett. Nuovo Cimento 1, 419 (1971)

[8] Gelfand I.M., Minlos R.A., Shapiro Z.Y., Representation of group of
rotations and group of Lorenth, Fizmatgiz, 1958.

[9] Lyubarskiy G.Y., Theory of group and its application in physics,
Fizmatgiz, 1957.


10. Al. L.Efros and M. Rosen Phys. Rev B58, 7120-7135 (1998)

11. F.M. Gashimzade, A.M. Babaev, M.A. Bagirov J.Phys.: Condens. Matter 12,
7923-7932 (2000)

\end{document}